\documentclass[abstract,headings=normal]{scrartcl}

\pdfoutput=1

\usepackage[automark]{scrpage2}
\usepackage[english]{babel}
\usepackage[T1]{fontenc}
\usepackage[applemac]{inputenc}
\usepackage{graphicx}
\usepackage{amsmath}
\usepackage{amssymb}
\usepackage{amsthm}
\usepackage{subfig}
\usepackage{hyperref}

\newcommand{\C}{\mathbb{C}}
\newcommand{\Z}{\mathbb{Z}}

\newcommand{\T}{\mathcal{T}}

\newcommand{\Hvier}{H\textsuperscript{4}}
\newcommand{\Hsechs}{H\textsuperscript{6}}

\newtheorem{theo}{Theorem}[section]
\newtheorem{lemma}[theo]{Lemma}
\theoremstyle{definition}
\newtheorem*{defi}{Definition}
\theoremstyle{remark}
\newtheorem*{rem}{Remark}

\graphicspath{{./graphics/}}
\DeclareGraphicsExtensions{.pdf}

\begin{document}

\title{On multidimensional consistent systems of asymmetric quad-equations}

\author{Raphael Boll\footnote{Institut f\"ur Mathematik, MA 7-2, Technische Universit\"at Berlin, Str.~des~17.~Juni~136, 10623 Berlin, Germany; e-mail: boll@math.tu-berlin.de}}

\maketitle
\begin{abstract} Multidimensional Consistency becomes more and more important in the theory of discrete integrable systems. Recently, we gave a classification of all 3D consistent 6-tuples of equations with the tetrahedron property, where several novel asymmetric systems have been found. In the present paper we discuss higher-dimensional consistency for 3D consistent systems coming up with this classification. In addition, we will give a classification of certain 4D consistent systems of quad-equations. The results of this paper allow for a proof of the Bianchi permutability among other applications.\par
\vspace{0.5cm}
\end{abstract}

\section{Introduction}
One of the definitions of integrability of discrete equations, which becomes increasingly popular in the recent years, is based on the notion of multidimensional consistency. For two-dimensional systems, this notion was clearly formulated first in \cite{NW}, and it was pushed forward as a synonym of integrability in \cite{quadgraphs,Nijhoff}. The outstanding importance of 3D consistency in the theory of discrete integrable systems became evident no later than with the appearance of the well-known ABS-classification of integrable equations in \cite{ABS1}. This classification deals with quad-equations which are set on the faces of an elementary cube in such a manner that all faces carry similar equations differing only by the parameter values assigned to the edges of a cube. Moreover, each single equation admits a $D_4$ symmetry group. Thus, ABS-equations from \cite{ABS1} can be extended in a straightforward manner to the whole of $\Z^{m}$.  In \cite{ABS2}, a relaxed definition of 3D consistency was introduced: the faces of an elementary cube are allowed to carry a priori different quad-equations. The classification performed in \cite{ABS2} is restricted to so-called equations of type~Q, i.e., those equations whose biquadratics are all non-degenerate (a precise definition will be recalled in Section \ref{faces}). In \cite{Atk1} and \cite{Todapaper}, numerous asymmetric systems of quad-equations have been studied, which also include equations of type H (with degenerate biquadratics). A classification of such systems has been given in  \cite{classification}.  This classification covers the majority of systems from \cite{Atk1} and all systems from \cite{Todapaper}, as well as equations from \cite{Hietarinta}, \cite{LY} and \cite{HV}. Moreover, it contains also a number of novel systems. The results of this local classification lead to integrable lattice systems via a procedure of reflecting the cubes in a suitable way (see \cite{classification} for details).
As shown in \cite{Atk1}, \cite{Todapaper},  \cite{classification}, the asymmetric systems still can be seen as families of B\"acklund transformations, and they lead to zero curvature representations of participating quad-equations. Lagrangian structures of this systems were discussed in \cite{lagrangian}.\par
The present paper is devoted to higher-dimensional (i.e.~four-dimensional) consistency of these systems. 4D consistency seems to be very important, because it allows for a proof of Bianchi permutability of the corresponding B\"acklund transformations (see \cite{bianchi,ABS1,classification}). Moreover, the present paper gives a confirmation that it makes sense to consider the Lagrangian structures from \cite{lagrangian} in a higher dimensional lattice. Furthermore we will give a classification of certain 4D consistent systems of quad-equations.\par
The outline of the paper is as follows: In Sections~\ref{faces} and \ref{cubes} we will introduce notations and recapitulate the classification results from \cite{classification} which are relevant for the present paper. Moreover, in Section~\ref{cubes} we will introduce a new object, the so-called \emph{super-consistent eight-tuples} on \emph{decorated 3D cubes} which can be seen as one of the main ideas of this paper. Then, in Section~\ref{consi} we will introduce the concept of 4D consistency and present a consistency criterion. Then, the topic of Section~\ref{embed} is the embedding in four-dimensional lattices. Finally, Section~\ref{classi} is devoted to the classification of 4D consistent systems.

\section{Quad-equations on a single square} \label{faces}

\begin{figure}[htbp]
   \centering
   \subfloat[Quad-equation of type Q]{\label{fig:3}\includegraphics[scale=1.2]{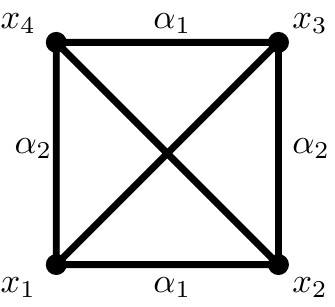}}\qquad
   \subfloat[Rhombic version of a quad-equation of type \Hvier]{\label{fig:2a}\includegraphics[scale=1.2]{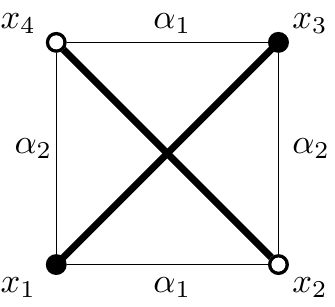}}\qquad
   \subfloat[Trapezoidal version of a quad-equation of type \Hvier]{\label{fig:2b}\includegraphics[scale=1.2]{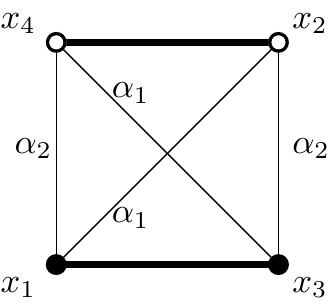}}
   \caption{Biquadratics patterns of quad-equations; non-degenerate biquadratics are indicated by thick lines}
\label{fig:2}
\end{figure}

We start with introducing relevant objects and notations. A \emph{quad-equation} is a relation of the type $Q\left(x_{1},x_{2},x_{3},x_{4}\right)=0$, where $Q\in\C\left[x_{1},x_{2},x_{3},x_{4}\right]$ is an irreducible multi-affine polynomial. It is convenient to visualize a quad-equation by an elementary square whose vertices carry the four fields $x_1,x_2,x_3,x_4$, cf.~Figure~\ref{fig:2}. For quad-equations without pre-supposed symmetries, the natural classification (see \cite{ABS2,classification}) problem is posed modulo the action of four independent M\"obius transformations of all four variables:
\[
Q(x_1,x_2,x_3,x_4)\; \rightsquigarrow\;
\prod_{k=1}^4(c_kx_k+d_k)\cdot Q\left(\frac{a_1x_1+b_1}{c_1x_1+d_1},\frac{a_2x_2+b_2}{c_2x_2+d_2},
\frac{a_3x_3+b_3}{c_3x_3+d_3},\frac{a_4x_4+b_4}{c_4x_4+d_4}\right).
\]
In the simplest situation, a quad-equation is thought of as an elementary building block of a discrete system $Q(x_{m,n},x_{m+1,n},x_{m+1,n+1},x_{m,n+1})=0$ for a function $x:\Z^2\to\C$, but we will also encounter more tricky ways of composing discrete systems on $\Z^2$ from different quad-equations.\par
A complete classification of multi-affine polynomials is given in \cite{classification}. Due to this classification quad-equations can be divided in three groups: \emph{Type~Q} equations with all six biquadratics are non-degenerate, \emph{type~\Hvier} equations with exactly two out of six biquadratics are non-degenerate and \emph{type~\Hsechs} equations with all biquadratics are degenerate. For more informations one can consult \cite{classification}.\par
For type~Q and type~\Hvier\ equations we introduced so-called \emph{biquadratics patterns} (see Figure~\ref{fig:2}) in \cite{lagrangian}. Here, non-degenerate biquadratics are indicated by thick lines. In the case of type~\Hvier\ equations there are two different possibilities of arranging the fields to the vertices of the square which we call \emph{rhombic} and \emph{trapezoidal version} of a quad-equation. In addition, in this case we have a black-and-white coloring of vertices which is coming from the fact that discriminants of type~\Hvier\ equations can belong to two non-equivalent classes. We mention here that non-degenerate biquadratics always connect vertices of the same color whereas the vertices of edges carrying degenerate biquadratics are always of different colors. For more informations we refer to \cite{classification,lagrangian}. The canonical forms $Q_{1}^{\epsilon}$ -- $Q_{3}^{\epsilon}$ (see \cite{lagrangian}) and $Q_{4}$ (see \cite{ABS2}) of type~Q equations and the canonical forms $H_{1}^{\epsilon}$ -- $H_{3}^{\epsilon}$ (see \cite{lagrangian}) depend on two parameters $\alpha_1,\alpha_2$ which can be assigned to the pairs of opposite edges of the elementary square.\par
\begin{figure}[htbp]
   \centering
   \includegraphics[scale=1]{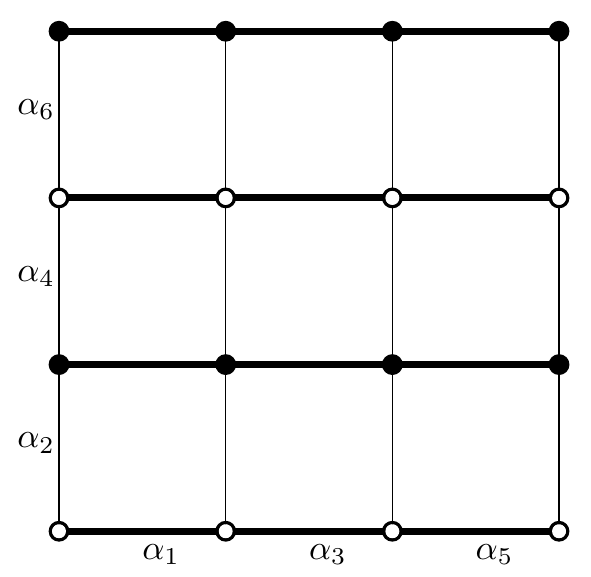}
   \caption{Integrable system on $\mathbb Z^2$ consisting of quad-equations $H_k^\epsilon$}
\label{fig:6}
\end{figure}
\begin{figure}[htbp]
   \centering
   \subfloat[Elementary cell of of the $\Z^{2}$-lattice]{\label{fig:6b}\includegraphics{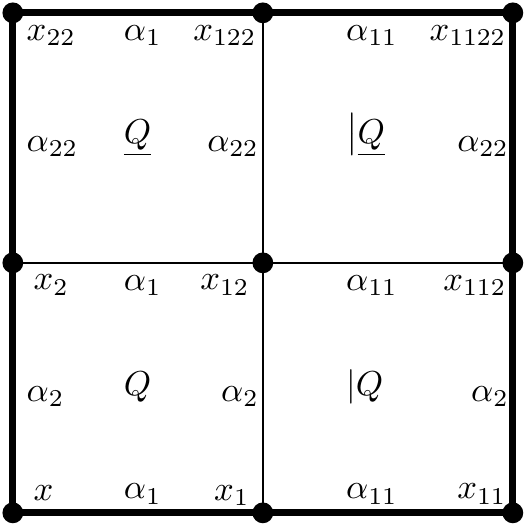}}\qquad
   \subfloat[Translation of elementary cells]{\label{fig:6a}\includegraphics{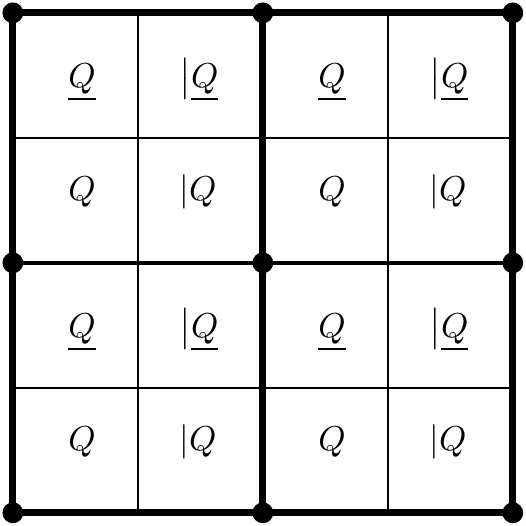}}
   \caption{Embedding in $\Z^{2}$}
\end{figure}

Concerning discrete systems composed of quad-equations, we mention that the rhombic version of any equation $H_k^\epsilon$ can be imposed on any bipartite quad-graph (cf.~also \cite{XP}). As for the trapezoidal version of $H_k^\epsilon$, it can be imposed on the regular square lattice $\Z^2$ with horizontal (or vertical) rows colored alternately black and white, see Figure~\ref{fig:6}.
In general, one can interpret the embedding procedure in the lattice $\Z^{2}$ by reflecting the squares (see \cite{classification}), i.e.~for a quad-equation
\[
Q\left(x,x_{1},x_{2},x_{12};\alpha_{1},\alpha_{2}\right)=0
\]
the three other quad-equations of an elementary cell are given by $\left|Q\right.=0$, $\underline{Q}=0$ and $\left|\underline{Q}\right.=0$ with
\begin{align*}
&\left|Q\right.\left(x,x_{1},x_{2},x_{12};\alpha_{1},\alpha_{2}\right):=Q\left(x_{1},x,x_{12},x_{2};\alpha_{1},\alpha_{2}\right),\\
&\underline{Q}\left(x,x_{1},x_{2},x_{12};\alpha_{1},\alpha_{2}\right):=Q\left(x_{2},x_{12},x,x_{1};\alpha_{1},\alpha_{2}\right)\quad \text{and}\\
&\left|\underline{Q}\right.\left(x,x_{1},x_{2},x_{12};\alpha_{1},\alpha_{2}\right):=Q\left(x_{12},x_{2},x_{1},x;\alpha_{1},\alpha_{2}\right).
\end{align*}
The arguments of the polynomials $\left|Q\right.$, $\underline{Q}$ and $\left|\underline{Q}\right.$ on the neighboring squares can be seen on Figure~\ref{fig:6b}.
The whole lattice $\Z^{2}$ can be filled by translating this elementary cell (see Figure~\ref{fig:6a}).
\begin{rem} In the case of equations of type \Hsechs, there are no parameters which can be assigned to the edges. Therefore, in this case we have to omit the parameters in the above considerations.
\end{rem}
Integrability of such a non-autonomous system of quad-equations understood as 3D consistency has been discussed in \cite{classification}.

\section{3D consistent six-tuples of quad-equations} \label{cubes}

\begin{figure}[htbp]
   \centering
   \subfloat[A 3D consistent six-tuple]{\label{fig:cube}\includegraphics{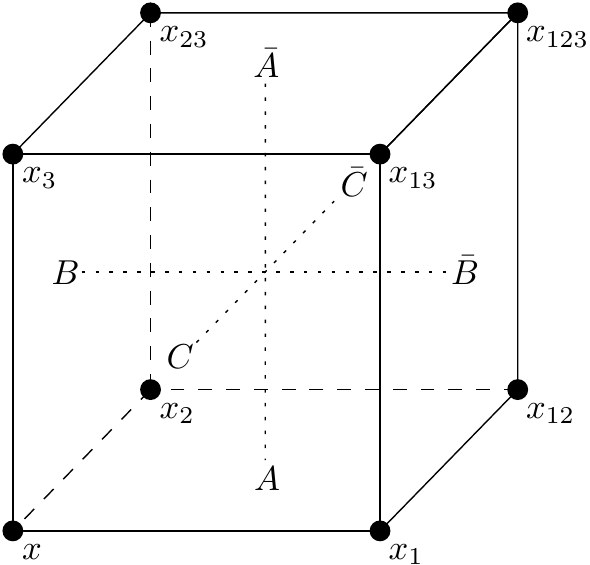}}\qquad
   \subfloat[Making tetrahedra to faces]{\label{fig:cube2}\includegraphics{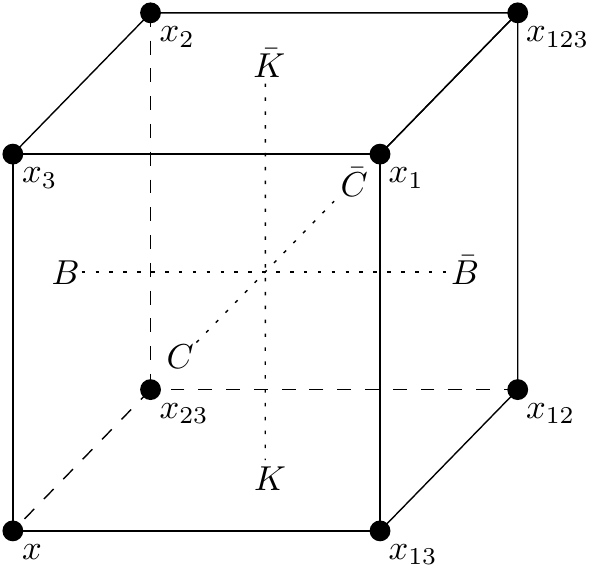}}
   \caption{Equations on a cube}
\end{figure}

\begin{figure}[htbp]
   \centering
   \subfloat[First case: all face equations of type \Hvier, tetrahedron equations of type Q]{\label{fig:5a}\includegraphics[scale=1.2]{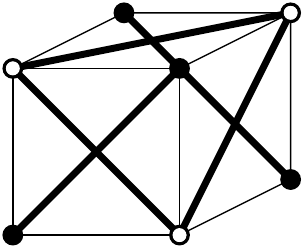}}\qquad
   \subfloat[Second case: two pairs of face equations and tetrahedron equations of type \Hvier, one pair of face equations of type Q]{\label{fig:5b}\includegraphics[scale=1.2]{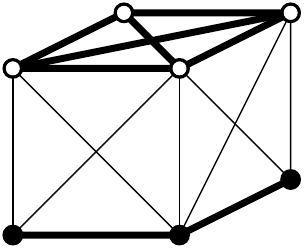}}\qquad
   \subfloat[Third case: all face and tetrahedron equations of type \Hvier]{\label{fig:5c}\includegraphics[scale=1.2]{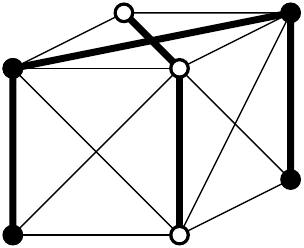}}
   \caption{Biquadratics patterns}
   \label{fig:5}
\end{figure}

We will now consider six-tuples of (a priori different) quad-equations assigned to the faces of a 3D cube:
\begin{align}\label{system}
&A\left(x,x_{1},x_{2},x_{12}\right)=0,&
&\bar{A}\left(x_{3},x_{13},x_{23},x_{123}\right)=0,\notag\\
&B\left(x,x_{2},x_{3},x_{23}\right)=0,&
&\bar{B}\left(x_{1},x_{12},x_{13},x_{123}\right)=0,\\
&C\left(x,x_{1},x_{3},x_{13}\right)=0,&
&\bar{C}\left(x_{2},x_{12},x_{23},x_{123}\right)=0,\notag
\end{align}
see Figure~\ref{fig:cube}. Such a six-tuple is \emph{3D consistent} if, for arbitrary initial data $x$, $x_{1}$, $x_{2}$ and $x_{3}$,
the three values for $x_{123}$ (calculated by using $\bar{A}=0$, $\bar{B}=0$ or $\bar{C}=0$) coincide. A 3D consistent six-tuple is said to possess the \emph{tetrahedron property} if there exist two polynomials $K$ and $\bar{K}$ such that the equations
\begin{align*}
&K\left(x,x_{12},x_{13},x_{23}\right)=0,&
&\bar{K}\left(x_{1},x_{2},x_{3},x_{123}\right)=0
\end{align*}
are satisfied for every solution of the six-tuple. It can be shown that the polynomials $K$ and $\bar{K}$ are multi-affine and irreducible (see \cite{classification}).
A complete classification of 3D consistent six-tuples~\eqref{system} possessing the tetrahedron property, modulo (possibly different) M\"obius transformations of the eight fields $x$, $x_{i}$, $x_{ij}$, $x_{123}$, is given in \cite{classification}. We mention one of the important general properties of such six-tuples:
\begin{lemma}\label{lem:opposite}
Equations on opposite faces are of the same type and have the same biquadratics pattern.
\end{lemma}
In the present article, we consider mostly six-tuples~\eqref{system} whose equations are not all of type~Q and which do not contain type~\Hsechs\ equations. Taking into account Lemma \ref{lem:opposite}, one easily sees that three different combinatorial arrangements of biquadratics patterns are possible, as indicated on Figure~\ref{fig:5}.
Arrangements on Figures~\ref{fig:5a} and \ref{fig:5b} are related to each other by the following construction, see \cite{classification} for a proof:
\begin{lemma} \label{tetrahedronUse}
Consider a 3D consistent six-tuple~\eqref{system} possessing the tetrahedron property expressed by the two equations
\begin{align*}
&K\left(x,x_{12},x_{13},x_{23}\right)=0,&
&\bar{K}\left(x_{1},x_{2},x_{3},x_{123}\right)=0.
\end{align*}
Then the six-tuple
\begin{align}\label{dualsystem}
&K\left(x,x_{12},x_{13},x_{23}\right)=0,&
&\bar{K}\left(x_{1},x_{2},x_{3},x_{123}\right)=0,\notag\\
&B\left(x,x_{2},x_{3},x_{23}\right)=0,&
&\bar{B}\left(x_{1},x_{12},x_{13},x_{123}\right)=0,\\
&C\left(x,x_{1},x_{3},x_{13}\right)=0,&
&\bar{C}\left(x_{2},x_{12},x_{23},x_{123}\right)=0,\notag
\end{align}
assigned to the faces of a 3D cube as on Figure~\ref{fig:cube2}, is 3D consistent and possesses the tetrahedron property expressed by the two equations
\begin{align*}
&A\left(x,x_{1},x_{2},x_{12}\right)=0,&
&\bar{A}\left(x_{3},x_{13},x_{23},x_{123}\right)=0.
\end{align*}
3D consistency of \eqref{dualsystem} is understood as the property of the initial value problem with the initial date $x$, $x_{3}$, $x_{13}$ and $x_{23}$.
\end{lemma}
The way from six-tuple~\eqref{system} to six-tuple~\eqref{dualsystem} can be described by flipping the assignment to the vertices of $x_{1}$ and $x_{13}$ and furthermore of $x_{2}$ and $x_{23}$. This lemma suggests the consideration of the eight-tuple consisting of six face equations and two tetrahedron equations as a separate object.
\begin{defi}
For a 3D consistent six-tuple of quad-equations possessing the tetrahedron property, the set consisting of the six face equations and the two tetrahedron equations is called a \emph{super-consistent eight-tuple} on a \emph{decorated cube}.
\end{defi}
The combinatorial structure behind a super-consistent eight-tuple, called a decorated cube in the above definition, consist of
\begin{itemize}
\item the eight vertices of a cube carrying the fields,
\item the twelve edges and twelve face diagonals carrying the parameters (if we consider every equation in this tuple is not of type \Hsechs) and
\item the six faces and two tetrahedra carrying the equations.
\end{itemize}
Thus, it is richer than the standard 3D cube. Nevertheless, we will illustrate a super-consistent eight-tuple as a 3D cube in our pictures.\par
\begin{figure}[htbp]
   \centering
   \includegraphics{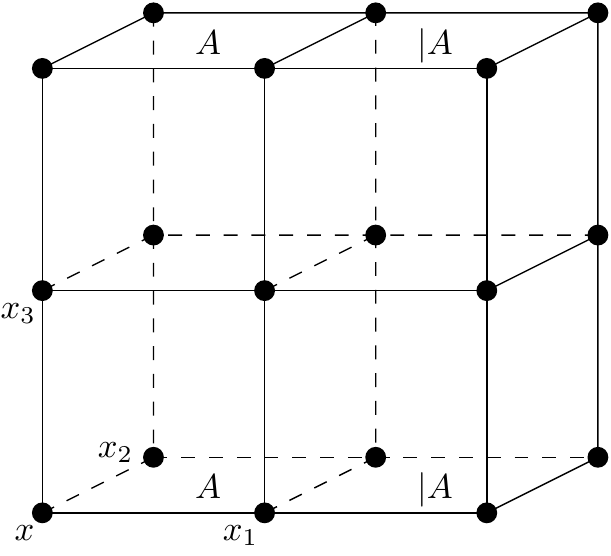} 
   \caption{Embedding procedure is independent of the order of reflections}
   \label{fig:16}
\end{figure}
All mentioned six-tuples can be embedded in the lattice $\Z^{3}$ by reflecting the cubes (see \cite{classification}). In addition to the considerations of the embedding in the lattice $\Z^{2}$ at the end of Section~\ref{faces} one has to think about the fact that the embedding procedure is independent of the order of reflections. To show this, we consider the situation demonstrated in Figure~\ref{fig:16}, where one can easily see that the chain of reflections from the polynomial $A$ on the bottom face to the polynomial $A$ on the top face, then from $A$ on the top face to $\left|A\right.$ on the top face, then from $\left|A\right.$ on the top face to $\left|A\right.$ on the bottom face and finally back to $A$ on the bottom face do not change the polynomial $A$ on the bottom face.

\section{4D Consistency and Consistency Criterion} \label{consi}
For the next two sections we return to the general situation, i.e.~we allow 3D consistent six-tuples containing type \Hsechs\ equations and 3D consistent systems containing only type~Q equations.\par
\begin{figure}[htbp]
   \centering
   \includegraphics{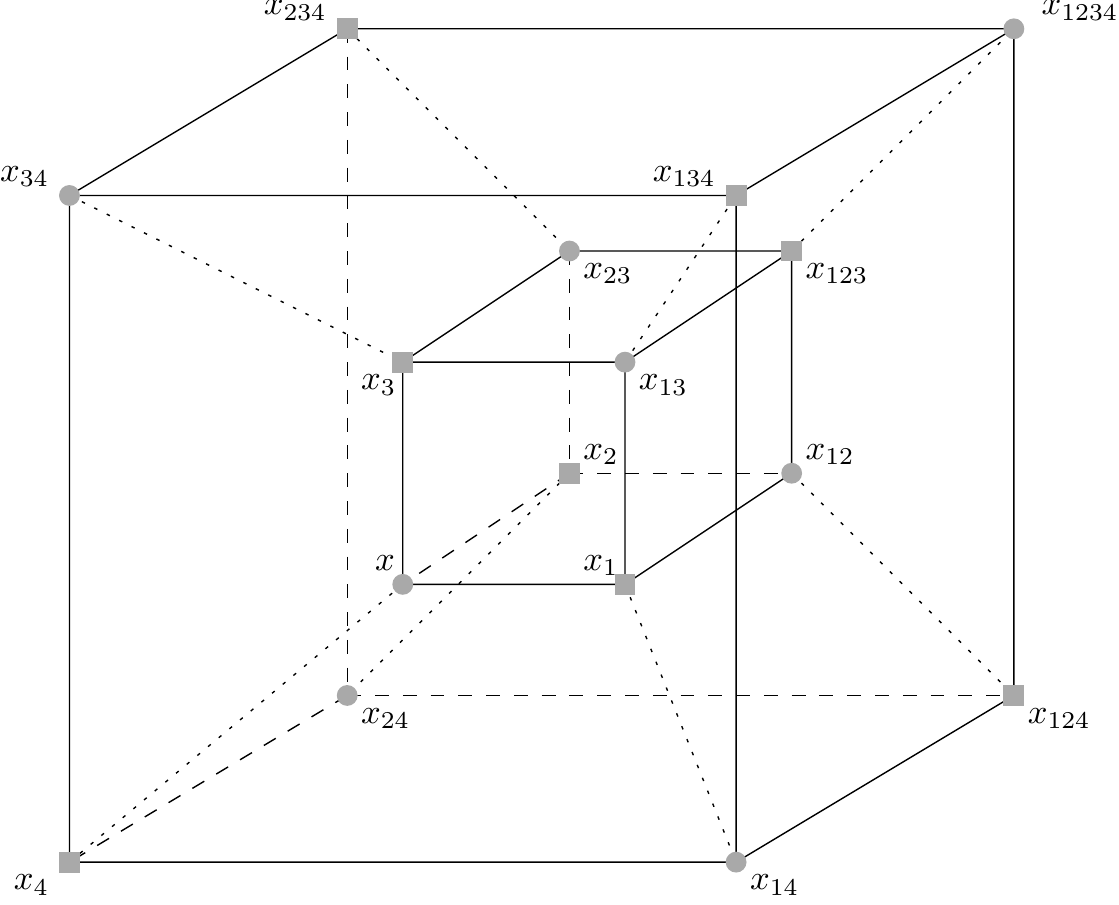} 
   \caption{24-tuple on 4D cube}
   \label{fig:4D cube}
\end{figure}
A 24-tuple of quad-equations on a 4D cube as shown in Figure~\ref{fig:4D cube} is \emph{4D consistent} if the six-tuples on all 3D facets are 3D consistent and the four values for $x_{1234}$ (calculated by using the six-tuples of 3D facets adjacent to $x_{1234}$) coincide for arbitrary initial data $x$, $x_{1}$, $x_{2}$, $x_{3}$ and $x_{4}$.\par
\begin{figure}[htbp]
   \centering
   \subfloat[Decorated 3D cube of vertices whose indices have an even number of digits]{\label{fig:a}\includegraphics{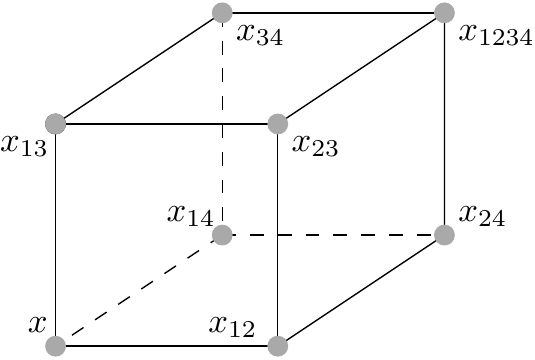}}\qquad
   \subfloat[Decorated 3D cube of vertices whose indices have an odd number of digits]{\label{fig:b}\includegraphics{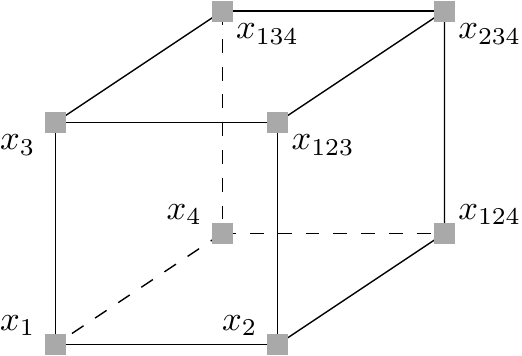}}
   \caption{super-consistent eight-tuples composed from tetrahedron equations of the 3D facets of a 4D consistent 24-tuple}
   \label{fig:tetras}
\end{figure}

We get the following consistency criterion:
\begin{theo} \label{cons} Consider a 24-tuple of quad-equations on a 4D cube and suppose that all eight six-tuples of equations corresponding to 3D facets are 3D consistent and possess the tetrahedron property. Then, the 24-tuple is 4D consistent if and only if the 16 tetrahedron equations build two super-consistent eight-tuples $\T$ and $\bar\T$ as demonstrated in Figure~\ref{fig:tetras}.
\begin{proof}
Consider the initial value problem with initial values $x$, $x_{1}$, $x_{2}$, $x_{3}$ and $x_{4}$. Then, we get unique values for $x_{12}$, $x_{13}$, $x_{14}$, $x_{23}$, $x_{24}$ and $x_{34}$ using the quad-equations on the corresponding faces. These values must satisfy the four tetrahedron equations $K\left(x,x_{ij},x_{ik},x_{jk}\right)=0$ ($\ell=1,2,3,4$ and $\left\{i,j,k\right\}=\left\{1,2,3,4\right\}\setminus\left\{\ell\right\}$) of the corresponding 3D facets. Assuming the tetrahedron equations $K\left(x_{i\ell},x_{j\ell},x_{k\ell},x_{1234}\right)=0$, we get four answers for $x_{1234}$. These are the eight equations of the eight-tuple demonstrated in Figure~\ref{fig:a}. If the 4D 24-tuple of quad-equations is consistent, the four answers must coincide and therefore the eight-tuple in Figure~\ref{fig:a} has to be super-consistent. On the other hand, if the eight-tuple in Figure~\ref{fig:a} is super-consistent, the four answers must also coincide and therefore the 4D 24-tuple has to be consistent, too. The super-consistency of the eight-tuple in Figure~\ref{fig:b} follows directly from the 4D consistency of the 24-tuple and the fact that the six-tuples on all 3D facets possess the tetrahedron property.\end{proof}\end{theo}

\section{Embedding in the 4D Lattice} \label{embed}
The next step is to extend a tuple of quad-equations on a 4D cube to system of quad-equations on a 4D lattice:
\begin{theo} A 4D consistent 24-tuple of quad-equations on a 4D cube as shown in Figure~\ref{fig:4D cube} can be extended to the whole lattice $\Z^{4}$ by reflecting the 4D cube, i.e., to get the neighboring 4D cube e.g.\ in the direction $4$ we take a copy of the first one reflect it in the 3D sub-space which is normal to direction $4$ and put it next to the original one.
\begin{proof}
While in the three-dimensional case, we have to reflect the 3D cube in two-dimensional planes, we have to reflect the 4D cube in three-dimensional sub-spaces. This means, e.g.~for the reflection in the 3D sub-space which is normal to the direction $4$ (we call this reflection in the direction $4$), that we switch the roles of fields in the quad-equations in the following way:
\begin{align*}
x&\leftrightarrow x_{4}, &x_{1}&\leftrightarrow x_{14}, &x_{2}&\leftrightarrow x_{24}, &x_{3}&\leftrightarrow x_{34},\\ x_{12}&\leftrightarrow x_{124}, &x_{13}&\leftrightarrow x_{134}, &x_{23}&\leftrightarrow x_{234}, &x_{123}&\leftrightarrow x_{1234}.
\end{align*}
\begin{figure}[htbp]
   \centering
   \includegraphics[scale=0.95]{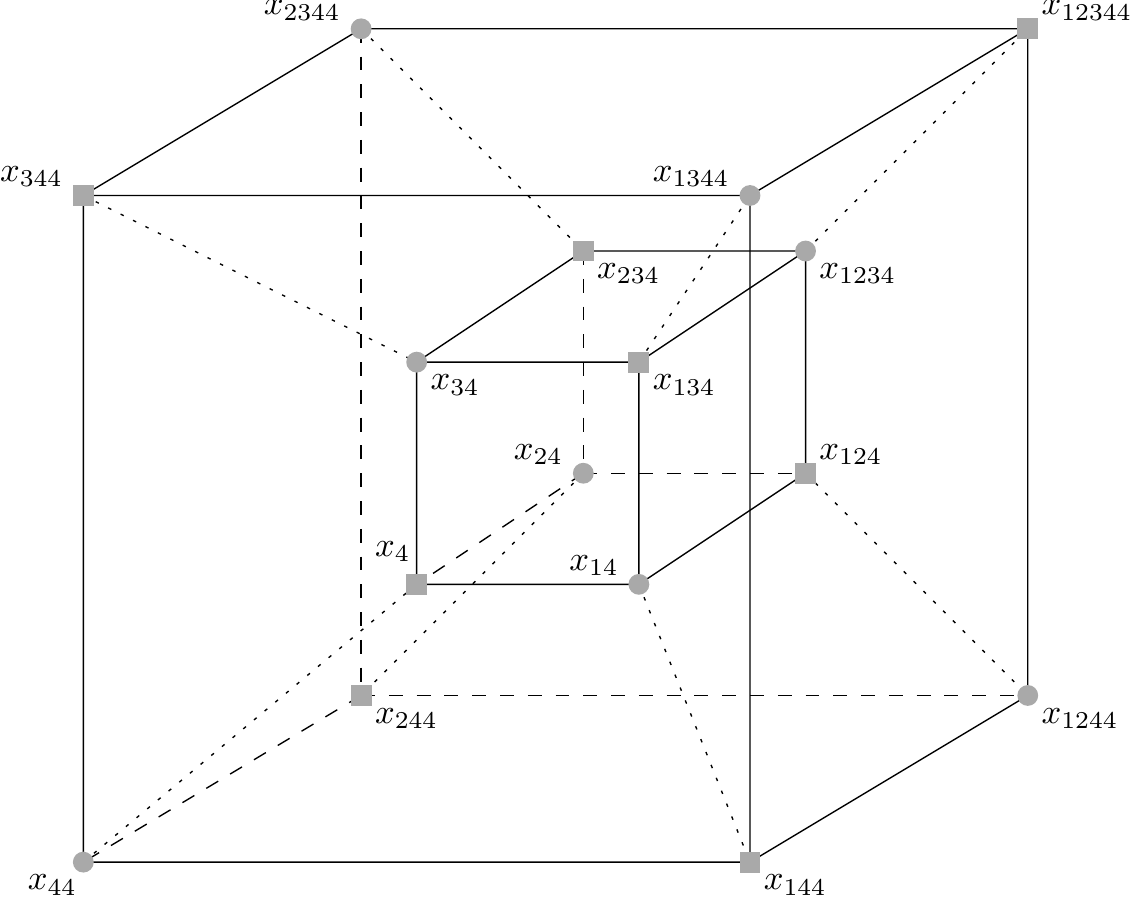} 
   \caption{24-tuple on the neighboring 4D cube in direction $4$ (shift of the cube in Figure~\ref{fig:4D cube} in the 4th coordinate direction)}
   \label{fig:4D cube1}
\end{figure}
\begin{figure}[htbp]
   \centering
   \includegraphics[scale=0.95]{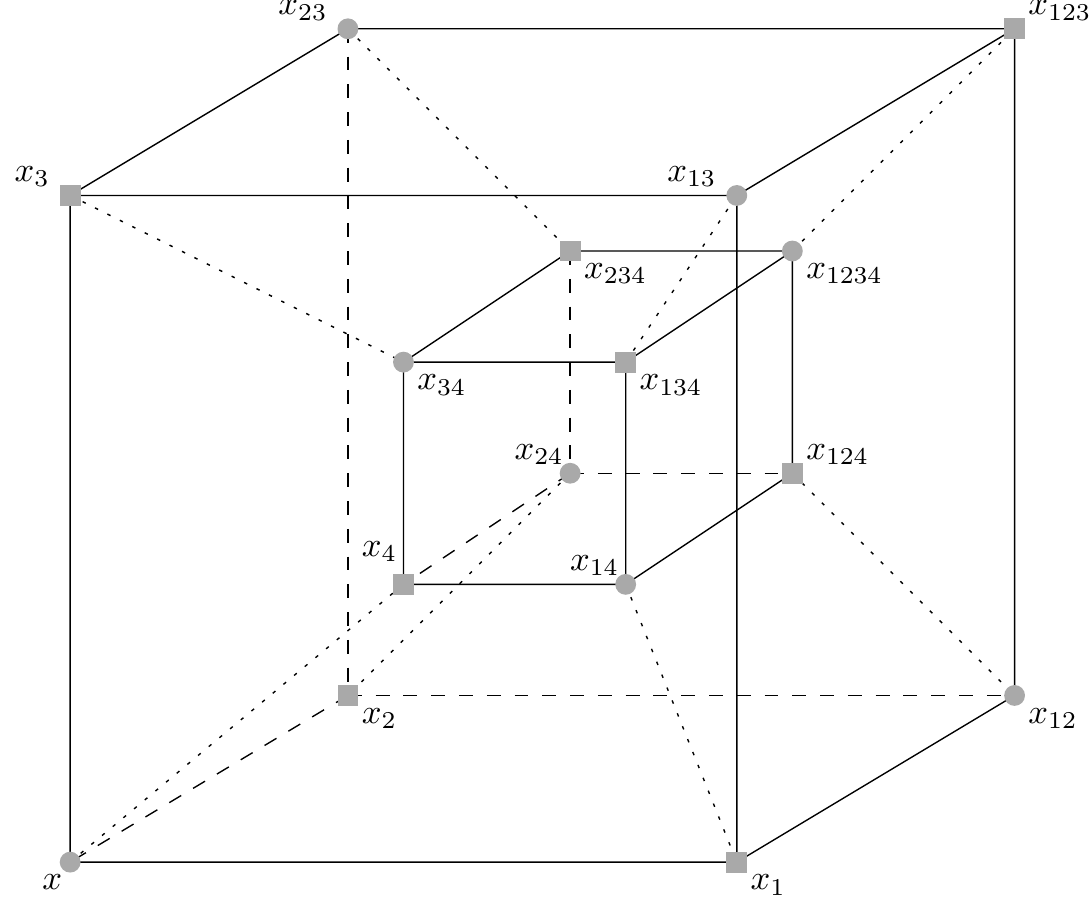} 
   \caption{24-tuple on the reflected 4D cube (reflection of the cube in Figure~\ref{fig:4D cube} in the hyperplane $x_{4}=\frac{1}{2}$: interchanges ``internal'' and ``external'' 3D facets)}
   \label{fig:4D cube2}
\end{figure}
This means that the neighboring cube demonstrated in Figure~\ref{fig:4D cube1} carries the same equations as the cube in Figure~\ref{fig:4D cube2}. The fields which are related to each other with quad-equations in the original system are the same as the ones related to each other in the reflected one. Indeed, considering only 3D facets, this reflection procedure reflects at most the sub-cube in planes parallel to faces of the sub-cube.\par
The above reflection procedure acts on the two decorated 3D cubes carrying the equations of the tetrahedron properties (see Figure~\ref{fig:tetras}) only by reflect them simultaneously in a plane which is parallel to its bottom faces. This fact can also be reached for the direction $1$ using Lemma~\ref{tetrahedronUse}. Therefore, due to Theorem~\ref{cons} the resulting system on the reflected 4D cube is consistent, too.\par
The last fact we have to prove is that reflecting first in the direction $3$ and then in the direction $4$ gives the same result as reflection first in the direction $4$ and then in the direction $3$. This can be easily verified by reflecting the corresponding decorated 3D cubes carrying the equations of the tetrahedron properties simultaneously.
This procedure of embedding in $\Z^{4}$ is compatible with the procedure of embedding in $\Z^{3}$ demonstrated above: If we fix, for instance, the third coordinate, i.e.,
\begin{align*} x&=x_{3},& x_{1}&=x_{13},& x_{2}&=x_{23}, & x_{4}&=x_{34},\\ x_{12}&=x_{123}, &x_{14}&=x_{134}, &x_{24}&=x_{234}, & x_{124}&=x_{1234},
\end{align*}
then the reflection in the direction $4$ can be described with
\begin{align*} x&\leftrightarrow x_{4}, &x_{1}&\leftrightarrow x_{14}, &x_{2}&\leftrightarrow x_{24}, & x_{12}&\leftrightarrow x_{124},
\end{align*}
which is exactly a reflection of the corresponding 3D facet.\end{proof}\end{theo}

\section{Classification of 4D Consistent 24-tuples of Quad-Equations} \label{classi}
Now we return again to the situation that all 3D facets have biquadratics patterns as mentioned in Figure~\ref{fig:5} or carry only equations of type~Q in order to give a classification of 4D consistent 24-tuples of quad-equations with the property that no face of the 4D cube carry a type \Hsechs\ equation.\par
Obviously, these 24-tuples allow for a black-and-white coloring of vertices consistent with the coloring of vertices for their 3D facets.\par
We have the following lemma:
\begin{lemma}
Every 4D consistent 24-tuple of quad-equations on a 4D cube containing no of type~\Hsechs\ with all six-tuples on its 3D facets possessing the tetrahedron property is completely determined by the two super-consistent eight-tuples $\T$ and $\bar{\T}$ consisting of the tetrahedron-equations of the 3D facets.
\begin{proof}
For all 2D faces of the 4D cube we know from $\T$ and $\bar{\T}$ the black-and-white coloring of vertices and for every biquadratic on a diagonal its parameter and if it is non-degenerate or degenerate. Due to the fact that vertices with the same color are always connected with edges carrying non-degenerate biquadratics and biquadratics on all other edges are degenerate we know in addition for all biquadratics on the edges of 3D facets if they are non-degenerate or degenerate. Therefore, we know the type and the biquadratics pattern for every equation of the 24-tuple. That is why the 24-tuple is completely determined since every field $x,x_{1},x_{2},\ldots,x_{1234}$ all takes part in at least one equation from $\T$ or $\bar{\T}$.\end{proof}
\end{lemma}
Moreover, consider two super-consistent eight-tuples $\T$ and $\bar{\T}$ on two decorated 3D cubes which contain only equations $Q_{k}^{\epsilon}$ and $H_{k}^{\epsilon}$ in the canonical forms. Suppose that their biquadratics patterns and black-and-white coloring correspond to those for tetrahedron equations of 3D facets of a 4D cube. Then one can easily reconstruct the corresponding 4D consistent 24-tuple by just looking at the given biquadratics.\par
Therefore, the only thing which remains to do in order to classify 4D consistent 24-tuples of quad-equations is to classify pairs of 3D consistent six-tuples with black-and-white coloring satisfying the above properties. This can be done by only using combinatorial arguments.\par
\begin{figure}[htbp]
   \centering
   \subfloat[]{\label{fig:tetra1.1}\includegraphics{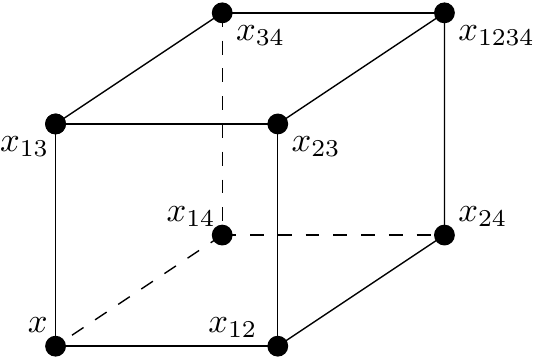}}\qquad\subfloat[]{\label{fig:tetra1.2}\includegraphics{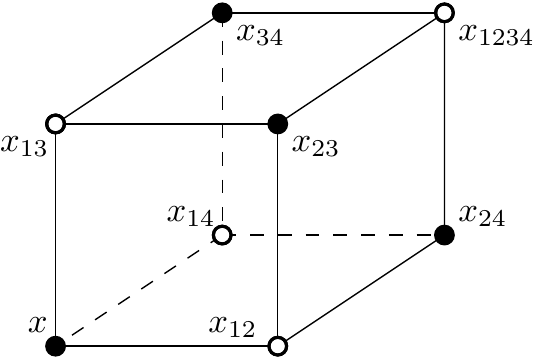}}\\\subfloat[]{\label{fig:tetra1.3}\includegraphics{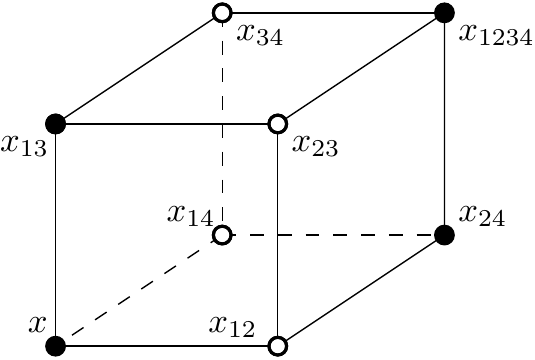}}
   \caption{Three possibilities for $\T$}
   \label{fig:tetra1}
\end{figure}

First we only consider the 3D cube $\T$ containing the eight vertices $x$, $x_{ij}$ and $x_{1234}$. One can easily see that the change of directions of the 4D consistent 24-tuple, i.e.~for example the following change of variables
\begin{align*} x_{1}&\leftrightarrow x_{2}, &x_{13}&\leftrightarrow x_{23}, &x_{14}&\leftrightarrow x_{24}, &x_{134}&\leftrightarrow x_{234},\end{align*}
acts by either reflections of $\T$ or the flipping procedure described in Lemma~\ref{tetrahedronUse}. Therefore, we only have to consider the three cases demonstrated in Figure~\ref{fig:tetra1}.\par
\begin{figure}[htbp]
   \centering
   \includegraphics{tetra11}\qquad \includegraphics{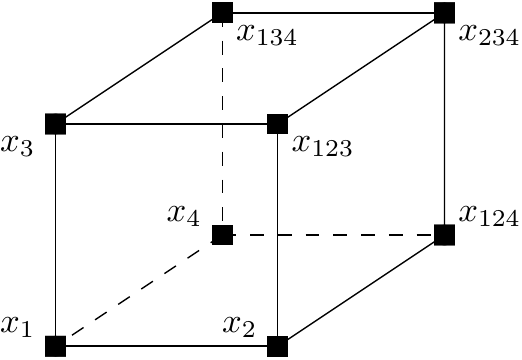}
   \caption{All 24 equations of the 4D tuple are of type Q}
   \label{fig:tetra2}
\end{figure}

\begin{figure}[htbp]
   \centering
   \includegraphics{tetra11}\qquad \includegraphics{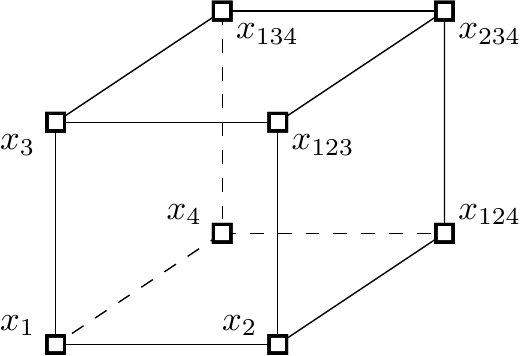}
   \caption{All 24 equations of the 4D tuple are of type~\Hvier\ in the rhombic version. All biquadratics patterns of the 3D facets are like in Figure~\ref{fig:5a}.}
   \label{fig:tetra3}
\end{figure}

We start with the cube in Figure~\ref{fig:tetra1.1}. The edge $\left(x,x_{1}\right)$ can be either non-degenerate, i.e.~$x_{1}$ must be a black vertex, or degenerate. Then, $x_{1}$ is a white vertex. Due to the fact that biquadratics on opposite edges are either both non-degenerate or both degenerate, the coloring of the other vertices of the cube $\bar{\T}$ with vertices $x_{i},x_{ijk}$ follows immediately in a unique manner and we get the two pairs demonstrated in Figures~\ref{fig:tetra2} and \ref{fig:tetra3}.\par
\begin{figure}[htbp]
   \centering
   \includegraphics{tetra5}\qquad \includegraphics{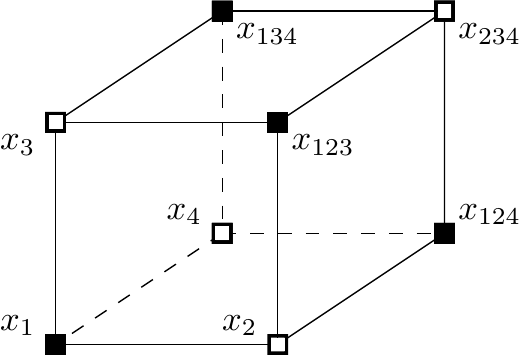}
   \caption{Twelve of the 24 equations of the 4D tuple are of type~\Hvier\ in the rhombic version, twelve in the trapezoidal version. Two of the biquadratics patterns of 3D facets are like in Figure~\ref{fig:5a}, six like in Figure~\ref{fig:5c}.}
   \label{fig:tetra4}
\end{figure}

\begin{figure}[htbp]
   \centering
   \includegraphics{tetra5}\qquad \includegraphics{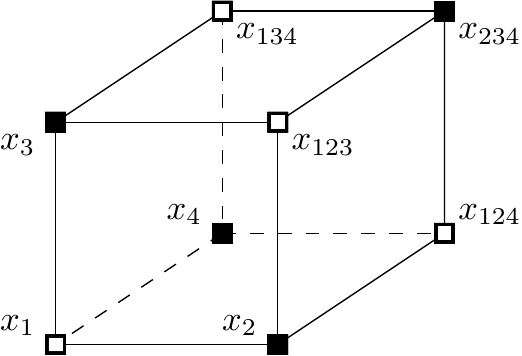}
   \caption{Twelve equations are of type~Q, twelve of type--\Hvier\ in the rhombic version. 2 3D facets carry only type~Q-equations, the biquadratics patterns of the other 3D facets are like in Figure~\ref{fig:5b}.}
   \label{fig:tetra5}
\end{figure}

Considering now the cube in Figure~\ref{fig:tetra1.2}, we have a similar situation: Again, the edge $\left(x,x_{1}\right)$ can be either non-degenerate, i.e.~$x_{1}$ must be a black vertex, or degenerate. Then, $x_{1}$ is a white vertex. Due to the fact that biquadratics on opposite edges are either both non-degenerate or both degenerate, the coloring of the other vertices of the cube $\bar{\T}$ with vertices $x_{i},x_{ijk}$ follows immediately in a unique manner and we get the two pairs demonstrated in Figures~\ref{fig:tetra4} and \ref{fig:tetra5}.\par
\begin{figure}[htbp]
   \centering
   \includegraphics{tetra7}\qquad \includegraphics{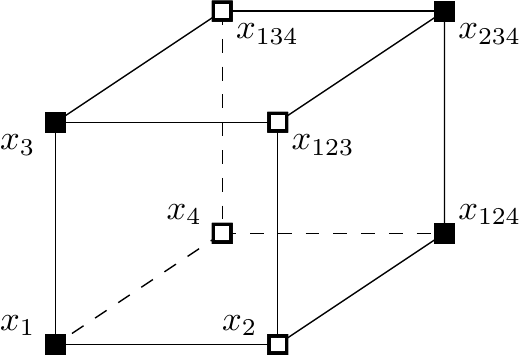}
   \caption{Four equations are of type~Q, four of type~\Hvier\ in the rhombic version, 16 in the trapezoidal version. Four biquadratics patterns are like in Figure~\ref{fig:5b}, four like in Figure~\ref{fig:5c}.}
   \label{fig:tetra6}
\end{figure}

\begin{figure}[htbp]
   \centering
   \includegraphics{tetra7}\qquad \includegraphics{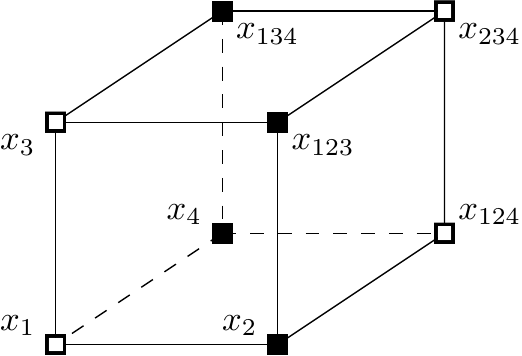}
   \caption{Four equations are of type~Q, four of type~\Hvier\ in the rhombic version, 16 in the trapezoidal version. Four biquadratics patterns are like in Figure~\ref{fig:5b}, four like in Figure~\ref{fig:5c}.}
   \label{fig:tetra7}
\end{figure}

In the same way, we get last two pairs (see Figures~\ref{fig:tetra6} and \ref{fig:tetra7}) considering the cube in Figure~\ref{fig:tetra1.3}.\par
The last to pairs are related to each other by the changes of directions $1\leftrightarrow4$ and $2\leftrightarrow3$, i.e.~the changes of variables
\begin{align*} x_{1}&\leftrightarrow x_{4}, &x_{2}&\leftrightarrow x_{3}, &x_{12}&\leftrightarrow x_{34}, &x_{13}&\leftrightarrow x_{24},\\&& x_{123}&\leftrightarrow x_{234}, &x_{124}&\leftrightarrow x_{134}.\end{align*}
Therefore, they are equivalent in the sense of our classification.\par
Summarizing the results of this section we can formulate the following theorem:
\begin{theo} Modulo independent M\"obius transformations of all fields, there are 16 different 4D consistent 24-tuples containing equations of type~Q and of type~\Hvier\ but no equations of type~\Hsechs\ and with all six-tuples on 3D facets are 3D consistent and possess the tetrahedron property. 15 of them correspond to the structures of Figures~\ref{fig:tetra2} -- \ref{fig:tetra6} with three possibilities for each structure enumerated by the index $k=1,2,3$ of the equations $H_{k}^{\epsilon},Q_{k}^{\epsilon}$. The remaining 16th 24-tuple contains only equations of the form $Q_{4}$ (see \cite{ABS1}).
\end{theo}
Finally, we mention that one can also generalize the Lemma~\ref{tetrahedronUse} to the four dimensional situation. The flipping procedure between the tuples described in Figures~\ref{fig:tetra3} and \ref{fig:tetra5} is given by the flips of the assignment to the vertices of the following pairs: $\left(x_{3},x_{13}\right)$, $\left(x_{2},x_{12}\right)$, $\left(x_{4},x_{14}\right)$ and $\left(x_{234},x_{1234}\right)$. The flipping procedure between the tuples in Figures~\ref{fig:tetra4} and \ref{fig:tetra7} is given by the flips of assignment to the vertices of the following pairs: $(x_{1},x_{13})$, $(x_{2},x_{23})$, $(x_{4},x_{34})$ and $(x_{123},x_{1234})$. One can easily see that there are no other such relations between different tuples, because of the different number of type~Q-equations.

\section{Conclusion}
Section~\ref{consi} give an idea on how to extend 3D consistent six-tuples of asymmetric quad-equations to a higher-dimensional cube which is not obvious from the outset, whereas Section~\ref{embed} describes how to embed this higher-dimensional cubes into higher-dimensional lattices. Together with Section~\ref{classi} a classification of 4D consistent systems of type~Q and type~\Hvier\ equations is now obtained which allows, among other applications, the proof of Bianchi-permutability (see \cite{bianchi}) for the corresponding B\"acklund transformations of all type~Q and type~\Hvier\ equations. Moreover, the present paper gives a confirmation that it makes sense to consider the Lagrangian structures from [BS11] in a higher dimensional lattice. The main idea which allowed us to arrive at all this results is the introduction of the new object of super-consistent eight-tuples on decorated 3D cubes.

\section*{Acknowledgments}
The author is supported by the Berlin Mathematical School and is indebted to Yuri~B. Suris for his continued guidance.

\bibliographystyle{amsalpha}
\bibliography{Quellen}

\end{document}